\documentclass[aip]{revtex4-1}

\usepackage{amsmath}
\usepackage{bm}
\usepackage{multirow}
\usepackage{svg}
\usepackage{soul}
\usepackage{color}

\newcommand{\bra}[1]{\ensuremath{\langle \: #1 \: |}}
\newcommand{\ket}[1]{\ensuremath{| \: #1 \: \rangle}}
\newcommand{\OvI}[2]{\ensuremath{\langle \: #1 \mid #2 \: \rangle}}

\newcommand{\quotes}[1]{``#1''}

\begin{document}

\title{Transcorrelated Density Matrix Renormalization Group}
\author{Alberto Baiardi}
\email{alberto.baiardi@phys.chem.ethz.ch}
\author{Markus Reiher}
\email{markus.reiher@phys.chem.ethz.ch}
\affiliation{ETH Z\"{u}rich, Laboratorium f\"{u}r Physikalische Chemie, Vladimir-Prelog-Weg 2, 8093 Z\"{u}rich, Switzerland.}
\date{September 30, 2020}

\begin{abstract}
We introduce the transcorrelated Density Matrix Renormalization Group (tcDMRG) theory for the efficient approximation
of the energy for strongly correlated systems.
tcDMRG encodes the wave function as a product of a fixed Jastrow or Gutzwiller correlator and a matrix product state.
The latter is optimized by applying the imaginary-time variant of time-dependent (TD) DMRG to the non-Hermitian transcorrelated Hamiltonian.
We demonstrate the efficiency of tcDMRG at the example of the two-dimensional Fermi-Hubbard Hamiltonian, a notoriously difficult target for the DMRG algorithm, for different sizes, occupation numbers, and interaction strengths.
We demonstrate fast energy convergence of tcDMRG, which indicates that tcDMRG could increase the efficiency of standard DMRG 
beyond quasi-monodimensional systems and provides a generally powerful approach toward the dynamic correlation
problem of DMRG.
\end{abstract}

\maketitle

\section{Introduction}
\label{sec:Intro}

Recent years have witnessed a growing interest in efficient configuration interaction (CI) based algorithms, such as full-CI Quantum Monte Carlo (FCIQMC) algorithm,\cite{Alavi2009_FCIQMCOriginal,Alavi2010_Initiator} different formulations of selected CI,\cite{Malrieu1973_CIPSI-Original,Malrieu1983_ImprovedCIPSI,Evangelista2016_ProjectorCI,HeadGordon2016_SelectedCI,Eriksen2017_VirtualOrbitalSelectedCI} and tensor-network-based approaches, such as the density matrix renormalization group (DMRG).\cite{White1992_DMRGBasis,White1993_DMRGBasis,Legeza2008_Review,Chan2008_Review,Zgid2009_Review,Marti2010_Review-DMRG,Schollwoeck2011_Review-DMRG,Chan2011,Wouters2013_Review,Kurashige2014_Review,Olivares2015_DMRGInPractice,Szalay2015_Review,Yanai2015,Knecht2016_Chimia,Baiardi2020_Review}
These advances allow for the calculation of full-CI (and complete active space-CI) energies for Hamiltonians with up to about 100 orbitals,\cite{Williams2020_ManyBodyMethods-Molecules,Eriksen2020_Benzene}, thereby extending the range of methods aiming at an accurate treatment of static correlation effects.
However, to develop efficient and reliable options for assessing the then still missing dynamical correlation effects 
remains to be a major challenge.
Approaches based on perturbation theory lead to a steep increase of the computational cost of both tensor-network-based approaches\cite{Kurashige2011_DMRGPT2,Sharma2014_Hylleraas-DMRG,Kurashige2014_DMRG-CumulantExpansion,Roemelt2016_DMRGPT2,Ren2016_InnerSpacePT,Wouters2016_DMRG-CASPT2,Chan2016_DMRG-NEVPT2,Sharma2016_QuasiDegenerate-PT,Freitag2017_DMRG-NEVPT2,Sharma2017_MRPT-DMRG,Guo2018_StochasticPT-DMRG,Guo2018_StochasticPT-DMRG}
and of selected CI algorithms\cite{Sharma2017_HBCI-Semistochastic,Loos2017_StochasticPT,Shrma2019_MRCI-WithoutRDM} due to the large size of the virtual orbital space and the unfavorable scaling of the number of elements of higher-order reduced density matrices.
Alternative strategies, such as the combination of such methods with density functional theory (DFT),\cite{Toulouse2004_srDFT-Theory,Olsen2014_PairDFT-Original,Hedegard2015_DMRG-srDFT,Giner2018_srDFT-BasisSet,Gagliardi2018_pDFT-DMRG} have been explored. However, they depend on the choice 
of a density functional and their ultimate accuracy has not been well established yet.

An alternative solution to the dynamical correlation problem is provided by explicitly-correlated methods\cite{Tenno2004_ExplicitlyCorrelated,Klopper2006_F12-Review,Valeev2012_R12-Review,Hattig2012_ExplicitlyCorrelated-Review}
that add to the wave function parametrization terms depending explicitly on inter-electronic distances. 
In this way, the energy convergence with basis set size is faster, because of better accounting for the consequences of the
singular Coulomb interactions of the electrons at short range. As a consequence, accurate results are obtained with smaller orbital spaces.
As a side remark, we note that the handling of short-range dynamic correlation by short-range density functionals has also
shown to regularize active orbital spaces making them more compact and stable with respect to changes in the active space.\cite{Hedegard2015_DMRG-srDFT}
So-called F12-based algorithms are now routinely applied in single-reference theories such as M{\o}ller-Plesset perturbation theory and coupled cluster theory,\cite{Kohn2010_ExplicitlyCorrelatedCC,Werner2018_ExplicitlyCorrelatedCC}
but their multi-reference generalizations are much less explored.\cite{Shiozaki2010_F12-PT2,Shiozaki2013_ExplicitlyCorrelated}

An alternative is the transcorrelation approach originally introduced by Boys and Handy.\cite{Handy1968_TranscorrelatedHamiltonian,Handy1971_TransCorrelated}
Transcorrelated methods parametrize the wave function as a product of a CI-like wave function and a fixed Jastrow factor.\cite{Jastrow1955_Original}
The latter is revolved from the wavefunction to the definition of the Hamiltonian by similarity transformation.
The former can then be optimized by applying standard quantum-chemical methods (such as Davidson subspace diagonalization)
to the resulting similarity-transformed Hamiltonian known as the transcorrelated Hamiltonian.
However, two factors have impeded so-far a widespread use of these approaches to quantum chemistry.
First, the transcorrelated Hamiltonian contains three-body interactions, which are technically hard
to include in common quantum-chemical methods as they require the implementation of new integrals over Gaussian basis functions.
Second, the transcorrelated Hamiltonian is a non-Hermitian operator, which prevents a trivial application of any variational method.
Even though several strategies have been proposed to overcome this second problem,\cite{Yanai2006_CanonicalTransformation,Luo2010_Transcorrelated-Optimization,Luo2010_VariationalTranscorrelated} none has proven to be a reliable alternative to F12-based schemes.

Luo and Alavi showed\cite{Alavi2018_FCIQMC-Transcorrelated} that methods based on imaginary-time evolution, such as FCIQMC, can be straightforwardly applied to non-Hermitian Hamiltonians.
Based on this idea, they showed that the convergence of FCIQMC is much faster when applied to transcorrelated Hamiltonians, both for the Fermi-Hubbard\cite{Alavi2019_RepulsiveHubbard-Transcorrelated} and for the non-relativistic electronic Hamiltonians.\cite{Alavi2019_Transcorrelated-Molecules}
The same idea has been recently exploited in the design of quantum-computing algorithms.\cite{Motta2020_Transcorrelated-QC,Tew2020_Transcorrelated-QC}
Inspired by the success of the transcorrelated FCIQMC theory, we introduce here the transcorrelated DMRG (tcDMRG). 
tcDMRG encodes the eigenvector of the transcorrelated Hamiltonian as a matrix product state (MPS) and optimizes it with imaginary-time time-dependent DMRG.
Among the various TD-DMRG theories proposed in the literature,\cite{Paeckel2019_Review} we rely in the present work on the tangent-space formulation of TD-DMRG\cite{Lubich2014_TimeIntegrationTT,Haegeman2016_MPO-TDDMRG,Baiardi2019_TDDMRG} that can support arbitrarily complex Hamiltonians, such as the transcorrelated one.

We apply tcDMRG to the calculation of the ground-state energy of the transcorrelated two-dimensional Fermi-Hubbard Hamiltonian that has a closed-form analytic expression and is therefore the ideal first test-case for tcDMRG.
We show that the energy convergence of tcDMRG is much faster than for standard DMRG in both weak and strong correlation regimes.
This suggests that the long-ranged interactions that make DMRG inefficient can be effectively reduced by similarity transformation. 
The eigenvector of the resulting non-Hermitian Hamiltonian can be effectively represented as a low-entanglement wave function and optimized with imaginary-time TD-DMRG.

\section{Transcorrelated DMRG theory}
\label{sec:TC_theory}

\subsection{Two-dimensional Fermi-Hubbard model}

The Fermi-Hubbard (FH) Hamiltonian for a two-dimensional lattice of width $W$, height $H$, $N_\alpha$ spin-up and $N_\beta$ spin-down electrons reads:

\begin{equation}
  \mathcal{H}_\text{FH}^\text{(r)} = 
    - t \sum_{\langle \bm{i}, \bm{j} \rangle} \sum_{\sigma} a_{\bm{i},\sigma}^\dagger a_{\bm{j},\sigma}
    + U \sum_{\bm{i}} n_{\alpha, \bm{i}} n_{\beta, \bm{i}} \, ,
  \label{eq:FH_Hamiltonian}
\end{equation}
where $\bm{i} = \left( i_x, i_y \right)$, with $0 \leq i_x \le H$ and $0 \leq i_y \le W$, and $\langle \bm{i}, \bm{j} \rangle$ denotes a sum over neighboring sites of the lattice. 
The first term is referred to as \quotes{hopping} term, while the second one is the \quotes{interaction} term, and the ratio between $t$ and $U$ defines the relative magnitude of the two terms. 
Moreover, $a_{\bm{i},\sigma}^\dagger$ is the creation operator for orbital $\bm{i}$ with spin $\sigma$, while $a_{\bm{i},\sigma}$ and $n_{\sigma,\bm{i}}$ are the corresponding annihilation and number operators, respectively.
The multireference character of the ground state of Eq.~(\ref{eq:FH_Hamiltonian}) increases with the $U/t$ ratio, and therefore large $U/t$ values correspond to a strongly correlated regime.

The Hamiltonian defined in Eq.~(\ref{eq:FH_Hamiltonian}) is usually known as the real-space ('r') representation of the Fermi-Hubbard Hamiltonian, since each orbital corresponds to a specific site of the lattice.
The real-space Hamiltonian of Eq.~(\ref{eq:FH_Hamiltonian}) can be expressed in a momentum-space ('m') representation (referred to as $\bm{k}$-space representation in the following) by a unitary transformation of the creation operator $a_{\bm{i}, \sigma}$ as follows:

\begin{equation}
  c_{\bm{k},\sigma} = \frac{1}{\sqrt{W \, L}} 
    \sum_{i_x=1}^{W} \sum_{i_y=1}^{H} e^{\mathrm{i} \bm{k} \cdot \bm{i}} a_{\bm{i},\sigma}
    \label{eq:SQOperator_Momentum}
\end{equation}
where $\bm{k} = \left( k_x, k_y \right)$ with $-W/2 \le k_x \leq W/2$ and $-H/2 \le k_y \leq H/2$.
The momentum-space representation of the Fermi-Hubbard Hamiltonian, obtained by combing Eqs.~(\ref{eq:SQOperator_Momentum}) and (\ref{eq:FH_Hamiltonian}), reads:

\begin{equation}
  \mathcal{H}_\text{FH}^\text{(m)} = 
    - t \sum_{\bm{k},\sigma} \epsilon_{\bm{k}} n_{\bm{k}, \sigma}
    + U \sum_{\bm{p},\bm{q},\bm{k},\sigma} 
      c_{\bm{p}-\bm{k},\sigma}^\dagger c_{\bm{q}+\bm{k},\bar{\sigma}}^\dagger
      c_{\bm{q},\bar{\sigma}} c_{\bm{p},\sigma}
  \label{eq:FH_MomentumSpace}
\end{equation}

Compared to Eq.~(\ref{eq:FH_Hamiltonian}), the hopping term has a simpler diagonal form, while the interaction term becomes more complex since it includes strings of four potentially different second-quantization operators.
Note that the Hamiltonian defined Eq.~(\ref{eq:FH_MomentumSpace}) has the same structure as the quantum chemical Hamiltonian in electronic structure theory.
The only major difference is that all interaction terms of Eq.~(\ref{eq:FH_MomentumSpace}) are scaled by the same factor (i.e., $U$), whereas such factor will be different for each combination of orbitals in the quantum chemical Hamiltonian.

It is known that, in the weakly correlated regime (small $U/t$), the ground state of the momentum-space Fermi-Hubbard Hamiltonian is mostly single-reference, whereas that of the real-space Hamiltonian is multi-reference.
For this reason, methods that approximate the ground-state wave function based on a single reference determinant, such as configuration interaction ones, are more efficient when applied to Eq.~(\ref{eq:FH_MomentumSpace}) than to Eq.~(\ref{eq:FH_Hamiltonian}).
However, in the strong correlation regime, the ground-state wave functions of both display a strong multi-reference character.
To tame such correlation effects, it is convenient to parametrize the ground-state wave function of the Fermi-Hubbard Hamiltonian with a Jastrow-like \textit{ansatz} as the following on:

\begin{equation}
  \ket{\Phi_\text{tc}} = e^{\tau} \ket{\Phi} \, ,
  \label{eq:JastrowParametrization}
\end{equation}
with

\begin{equation}
  \tau = J\sum_{\bm{l}} n_{\bm{l},\alpha} n_{\bm{l},\beta} \, .
  \label{eq:CorrelationFactor}
\end{equation}

$\ket{\Phi}$ is parametrized with standard quantum-chemical methods, such as Hartree-Fock (HF) or full-CI\cite{Scuseria2015_LieAlgebraic,Alavi2019_RepulsiveHubbard-Transcorrelated} (note that Eq.~(\ref{eq:JastrowParametrization}) applies to both Hamiltonian representations).
The Jastrow factor expressed in the second-quantization space is also known as Gutzwiller correlator, and is uniquely defined by the single parameter $J$.
Inspired by Ref.~\citenum{Alavi2019_RepulsiveHubbard-Transcorrelated}, where $\ket{\Phi}$ is parametrized as a full-CI wave function, we here encode $\ket{\Phi}$ as a matrix product state ($\Phi_\text{MPS}$).

The Jastrow parameter $J$ and the wave function $\ket{\Phi}$ can in principle be optimized simultaneously to minimize the energy functional variationally. 
However, for a given $J$ value, the optimal $\ket{\Phi}$ wave function is obtained as the right eigenfunction of the following similarity-transformed Hamiltonian, known as the transcorrelated Hamiltonian:

\begin{equation}
\mathcal{H}_\text{tcFH} \ket{\Phi_\text{tc}}
= \left( e^{-\tau} \mathcal{H}_\text{FH} e^{\tau} \right) \ket{\Phi_\text{tc}}
= E_\text{tc} \ket{\Phi_\text{tc}} \, ,
\label{eq:SimilarityTransformed}
\end{equation}

A closed-form expression for $\mathcal{H}_\text{tcFH}$ is obtained by evaluating $e^{-\tau} \mathcal{H}_\text{FH} e^{\tau}$ with a Baker-Campbell-Hausdorff formula.
This leads to a many-body expansion that, for the specific definition of the correlator given in Eq.~(\ref{eq:CorrelationFactor}), truncates at low order.
The transcorrelated real-space Fermi-Hubbard Hamiltonian reads

\begin{equation}
  \begin{aligned}
    \mathcal{H}_\text{tcFH}^\text{(r)} = 
      -&t \sum_{\langle \bm{i}, \bm{j} \rangle} \sum_{\sigma} a_{\bm{i},\sigma}^\dagger a_{\bm{j},\sigma}
      + U \sum_{\bm{i}} n_{\alpha,\bm{i}} n_{\beta,\bm{i}}
      +2t \sum_{\langle \bm{i}, \bm{j} \rangle} \sum_{\sigma} a_{\bm{i},\sigma}^\dagger a_{\bm{j},\sigma} 
      \cosh(J) n_{\bm{i},\bar{\sigma}} n_{\bm{j},\bar{\sigma}} \\
      -&t \left( e^J-1 \right) \sum_{\langle \bm{i}, \bm{j} \rangle} \sum_{\sigma} 
      a_{\bm{i},\sigma}^\dagger a_{\bm{j},\sigma} n_{\bm{j},\bar{\sigma}}
      - t \left( e^{-J}-1 \right) \sum_{\langle \bm{i}, \bm{j} \rangle} \sum_{\sigma}
      a_{\bm{i},\sigma}^\dagger a_{\bm{j},\sigma} n_{\bm{i},\bar{\sigma}} \, .
  \end{aligned}
  \label{eq:FH_Transcorrelated_Real}
\end{equation}

Eq.~(\ref{eq:FH_Transcorrelated_Real}) was first derived in Ref.~\citenum{Tsuneyuki2008_Transcorrelated-RealSpace}, and later applied to Hartree-Fock wave functions,\cite{Scuseria2015_LieAlgebraic} Monte Carlo-based methods,\cite{Neuscamman2011_NonStochastic-TC} and FCIQMC.\cite{Alavi2019_RepulsiveHubbard-Transcorrelated}
The momentum-space counterpart of Eq.~(\ref{eq:FH_Transcorrelated_Real}) was first derived in Ref.~\citenum{Alavi2019_RepulsiveHubbard-Transcorrelated} and reads:

\begin{equation}
\begin{aligned}
\mathcal{H}_\text{tcFH}^\text{(m)} 
= &- t \sum_{\bm{k},\sigma} \epsilon_{\bm{k}} n_{\bm{k}, \sigma}
+ \sum_{\bm{p},\bm{q},\bm{k},\sigma} 
\left( 
\frac{U}{2} - t \left[ (e^J-1) \epsilon_{\bm{p}-\bm{k}} 
+ (e^{-J}-1) \epsilon_{\bm{p}} \right]
\right)
c_{\bm{p}-\bm{k},\sigma}^\dagger c_{\bm{q}+\bm{k},\bar{\sigma}}^\dagger
c_{\bm{q},\bar{\sigma}} c_{\bm{p},\sigma} \\
& +2t \frac{\cosh(J)-1}{W^2 H^2} 
\sum_{\bm{p},\bm{q}, \bm{s}, \bm{k}, \bm{k}',\sigma} 
\epsilon_{\bm{p}-\bm{k}+\bm{k}'}
c_{\bm{p}-\bm{k},\sigma}^\dagger c_{\bm{q}+\bm{k}',\bar{\sigma}}^\dagger
c_{\bm{s}+\bm{k}-\bm{k}',\bar{\sigma}}^\dagger c_{\bm{s},\bar{\sigma}}
c_{\bm{q},\bar{\sigma}} c_{\bm{p},\sigma}
\end{aligned}
\label{eq:FH_Transcorrelated_Momentum}
\end{equation}

Correlation effects in the two-dimensional Fermi-Hubbard model are determined not only by the $U/t$ ratio, but also by the Hamiltonian boundary conditions. Following Ref.~\citenum{Alavi2019_RepulsiveHubbard-Transcorrelated}, we rely on full periodic boundary conditions, \textit{i.e.} the hopping term in Eq.~\ref{eq:FH_Hamiltonian} couples sites with $i_x=j_x$, $i_y$=0, $i_y=W$, and with $i_y=j_y$, $i_x$=0, and $i_x=H$.
Alternative choices are open boundary conditions, where no periodicity is imposed, and so-called cylindrical boundary conditions, where periodicity is imposed along only one of the two dimensions.\cite{Stoudenmire2012_2DHubbard-DMRG} A hybrid real/momentum space representation of the two-dimensional Fermi-Hubbard Hamiltonian\cite{Pollmann2016_Cylinder-DMRG,White2017_HybridSpace-DMRG} based on cylindrical boundary conditions can target lattices with up to $H=32$, $W=6$. 
Even though such sizes cannot be targeted with the setup adopted here, we rely on periodic boundary conditions to enhance correlation effects and, therefore, challenge the accuracy of tcDMRG.

\subsection{Imaginary-time DMRG optimization}

Both Eq.~(\ref{eq:FH_Transcorrelated_Real}) and Eq.~(\ref{eq:FH_Transcorrelated_Momentum}) define non-Hermitian operators, and this impedes a variational optimization of $E_\text{tc}$ with standard DMRG.
However, Alavi and co-workers\cite{Alavi2018_FCIQMC-Transcorrelated} proved that the right eigenvector of $\mathcal{H}_\text{tc}$, both in real- and momentum-space representation, can be optimized by imaginary-time evolution, \textit{i.e.}

\begin{equation}
\ket{\Phi_\text{tc}} = \lim_{t \rightarrow +\infty} e^{-\mathcal{H}_\text{tc} t} \ket{\Phi_\text{guess}} \, ,
\label{eq:ImaginaryTime}
\end{equation}
where $\ket{\Phi_\text{guess}}$ is a guess wave function such that  $\OvI{\Phi_\text{tc}}{\Phi_\text{guess}} \neq 0$.
Instead of evaluating the limit, Eq.~(\ref{eq:ImaginaryTime}) is often implemented by applying repeatedly the time-evolution operator $e^{-\mathcal{H}_\text{tc} \Delta t}$ for a finite time step $\Delta t$, until convergence.
Luo and Alavi solved Eq.~(\ref{eq:ImaginaryTime}) stochastically with FCIQMC,\cite{Alavi2018_FCIQMC-Transcorrelated} by representing $\ket{\Phi_\text{tc}}$ as an ensemble of discrete walkers. 
In the present work, we encode instead $\ket{\Phi_\text{tc}}$ as an MPS $\ket{\Phi_\text{MPS}^\text{tc}}$ and optimize it with the imaginary-time version of the density matrix renormalization group (DMRG) theory (iTD-DMRG).
The solution of Eq.~(\ref{eq:ImaginaryTime}) can be obtained by taking the $t \rightarrow + \infty$ limit of the solution of the following differential equation

\begin{equation}
\frac{\partial \ket{\Phi_\text{MPS}^\text{tc}(t)}}{\partial t} 
= - \mathcal{H}_\text{tc} \ket{\Phi_\text{MPS}^\text{tc}(t)} \, .
\label{eq:ImaginaryTime_TDSchro}
\end{equation}
where the wave function $\ket{\Phi_\text{MPS}^\text{tc}(t)}$ is expressed as an MPS,

\begin{equation}
\ket{\Phi_\text{MPS}^\text{tc}(t)} = \sum_{\bm{\sigma}} \sum_{\bm{m}} 
M_{1,m_1}^{\sigma_1}(t) M_{m_1,m_2}^{\sigma_2}(t) \cdots M_{m_{L-1},1}^{\sigma_L}(t)
\ket{\bm{\sigma}} \, .
\label{eq:MPS_Def}
\end{equation}

In an MPS, the CI tensor is replaced by a product of $L$ three-dimensional tensors, one per site $i$ ($M_{m_{i-1},m_i}^{\sigma_i}$). From a numeric analysis perspective, Eq.~(\ref{eq:MPS_Def}) is obtained from a standard CI expansion by replacing the CI tensor with its tensor-train factorization.
The index $\sigma_i$ in $ \bm{\sigma}= \left(\sigma_1,..,\sigma_L\right)$ (usually referred to as physical index) runs over all possible occupations for the $i$-th orbital (referred to as \quotes{site} in DMRG terminology).
The maximum dimension for the $m_{i-1}$ and $m_i$ indices is the \quotes{bond dimension} and tunes the accuracy of approximating a CI wave function as in Eq.~(\ref{eq:MPS_Def}).
DMRG will be efficient if an accurate representation of the wave function can be obtained with low $m$ values.
The area law\cite{Hastings2007_AreaLaw} ensures that this is the case for the ground-state wave function of short-ranged Hamiltonians.
In the DMRG context, \quotes{short-range} means that it exists a sorting of the orbitals such that only neighboring ones interact.
This is not the case of the two-dimensional Fermi-Hubbard model, neither in real-space, due to the off-diagonal hopping terms, nor in momentum-space, where the potential-energy is long-range.
However, as Alavi showed that the ground state of the Fermi-Hubbard Hamiltonian can be encoded as in Eq.~(\ref{eq:JastrowParametrization}) by expressing $\ket{\Phi}$ as a compact CI expansion, we aim here at showing that $\ket{\Phi}$ can be efficiently encoded as an MPS with a low bond dimension.

Eq.~(\ref{eq:ImaginaryTime_TDSchro}) cannot be solved exactly by fixing the bond dimension $m$ of the MPS at all times because the bond dimension of the MPS representation of $\mathcal{H} \Phi_\text{MPS}$ is larger than that of $\Phi_\text{MPS}$.\cite{Schollwoeck2011_Review-DMRG} 
Various TD-DMRG algorithms\cite{Vidal2004_TEBD,Feguin2006_Adaptive-TDDMRG,Ronca2017_TDDMRG-Targeting} approximate the solution to Eq.~(\ref{eq:ImaginaryTime_TDSchro}) with different strategies.
Here, we apply the so-called tangent-space approach\cite{Haegeman2016_MPO-TDDMRG} that replaces the imaginary-time time-dependent Schr\"{o}dinger equation by the following, projected counterpart

\begin{equation}
\frac{\partial \Phi_\text{MPS}(t)}{\partial t} = - \mathcal{P}_{\Phi_\text{MPS}(t)} \mathcal{H} \Phi_\text{MPS}(t) \, ,
\label{eq:Projected_TD}
\end{equation}
where $\mathcal{P}_{\Phi_\text{MPS}(t)}$ is the so-called tangent-space projector\cite{Lubich2014_TimeIntegrationTT,Haegeman2016_MPO-TDDMRG} which ensures that the bond dimension of the MPS remains constant during the propagation.
Lubich and co-workers\cite{Lubich2014_TimeIntegrationTT} derived the following closed-form expression for $\mathcal{P}_{\Phi_\text{MPS}(t)}$:

\begin{equation}
\mathcal{P}_{\Phi_\text{MPS}(t)} = \sum_{i=1}^{L-1} \left( 
\ket{a_i^{(l)} \sigma_i a_{i+1}^{(r)}} \bra{a_i^{(l)} \sigma_i a_{i+1}^{(r)}}
- \ket{a_{i+1}^{(l)} a_{i+1}^{(r)}} \bra{a_{i+1}^{(l)} a_{i+1}^{(r)}}
\right)
\label{eq:TangentSpaceProjector}
\end{equation}
where $\ket{a_i^{(l)}}$ is the left-renormalized basis for site $i$, defined recursively in terms of $ \ket{a_{i-1}^{(l)}}$ as:

\begin{equation}
\ket{a_i^{(l)}} = \sum_{a_{i-1} \sigma_i} 
M_{a_{i-1},a_i}^{\sigma_i} \ket{a_{i-1}^{(l)} \sigma_i} \, ,
\label{eq:LeftRenormalizedBasis}
\end{equation}
and $\ket{a_i^{(r)}}$ is defined analogously.
The differential equation obtained by combining Eqs.~(\ref{eq:Projected_TD}) and (\ref{eq:TangentSpaceProjector}) can be solved by approximating the resulting time-evolution operator with a second-order Trotter approximation.\cite{Haegeman2016_MPO-TDDMRG,Baiardi2019_TDDMRG}
Under these approximations, the MPS is propagated for a time step $\Delta t$ by updating the tensors $\bm{M}^{\sigma_i}$ one site after the other in a sweep-like fashion. 
For each site, the following differential equation is solved:

\begin{equation}
\frac{\mathrm{d} M_{a_{i-1},a_i}^{\sigma_i}}{\mathrm{d} t} 
= - \sum_{a_{i-1}',\sigma_i',a_i'} 
H_{a_{i-1}\sigma_i a_i, a_{i-1}'\sigma_i'a_i'} M_{a_{i-1}',a_i'}^{\sigma_i'} \, ,
\label{eq:LocalDifferentialEquation}
\end{equation}
where $H_{a_{i-1}\sigma_i a_i, a_{i-1}'\sigma_i'a_i'}$ is the representation of the Hamiltonian in the $\ket{a_i^{(l)} \sigma_i a_{i+1}^{(r)}}$ basis (referred to in the following as \quotes{site basis}).
Eq.~(\ref{eq:LocalDifferentialEquation}) is a linear differential equation that is solved with Lanczos-based algorithms.
The only approximation of our iTD-DMRG approach is therefore the Trotter factorization of the time-evolution operator.
This is a remarkable difference with other TD-DMRG formulations that support only short-ranged Hamiltonians\cite{Vidal2004_TEBD} or introduce additional approximations in the solution of the differential equation.\cite{Feguin2006_Adaptive-TDDMRG,Ronca2017_TDDMRG-Targeting,Frahm2019_TD-DMRG_Ultrafast}

For real-time evolutions, the propagation of the MPS requires an additional back-propagation step, associated to the second term of Eq.~(\ref{eq:TangentSpaceProjector}). Such step prevents that some components of the MPS are forward propagated twice.
However, as discussed by Haegeman and co-workers for spin lattices\cite{Haegeman2016_MPO-TDDMRG} and shown by us for vibrational Hamiltonians,\cite{Baiardi2019_TDDMRG} the back-propagation step can be neglected for imaginary-time evolution.

In the following, we will refer to imaginary-time DMRG optimization applied to transcorrelated Hamiltonian, either in real or in momentum space, as tcDMRG, and we will keep the iTD-DMRG acronym for imaginary-time optimization applied to non-transcorrelated, Hermitian Hamiltonians.

In conclusion, we highlight that the right eigenvector of the transcorrelated Hamiltonian can be in principle optimized with the non-Hermitian time-independent DMRG theory introduced by Chan and Van Voorhis in 2005\cite{Chan2005_DMRG-NonOrthogonal} that has been recently extended to classical statistical mechanics.\cite{Schollwock1999_NonHermitianDMRG,Helms2018_LargeDeviationFunction,Helms2020_DynamicalPhase}
However, iTD-DMRG is a more appealing optimization strategy, mainly for two reasons. 
First, the  diagonalization of the site Hamiltonian with iterative algorithms becomes challenging for non-Hermitian operators. This limited so far the bond dimension that can be targeted to 50-100. Conversely, the solution of the local differential equation of Eq.~\ref{eq:LocalDifferentialEquation} with iterative schemes\cite{Saad1992_MatrixExponential} is as complex for non-Hermitian operators as it is for Hermitian ones. Moreover, the non-Hermitian time-independent DMRG theory encodes both the left and right eigenfunctions of the Hamiltonian as MPSs. 
This is not the case of iTD-DMRG that parametrizes only the right eigenfunction. Alavi and co-workers showed\cite{Alavi2019_RepulsiveHubbard-Transcorrelated} that the CI representation of the right eigenfunction of the transcorrelated Hamiltonian is much more compact than that of the left one, and we will show in the next section that the same effect is observed in tcDMRG. For this reason, we expect the convergence of non-Hermitian TI-DMRG to be much slower than for tcDMRG due to the need of encoding both eigenfunctions as MPS.

\subsection{MPO representation of the transcorrelated Hamiltonian}

The representation of the Hamiltonian in a given site basis $\ket{a_i^{(l)} \sigma_i a_{i+1}^{(r)}}$, required to solve Eq.~(\ref{eq:LocalDifferentialEquation}), can be conveniently calculated by encoding $\mathcal{H}_\text{tc}$ as matrix product operator (MPO):\cite{Schollwoeck2011_Review-DMRG,Keller2015_MPS-MPO-SQHamiltonian,Chan2016_MPO-MPS}

\begin{equation}
\begin{aligned}
\mathcal{H} &= \sum_{\bm{\sigma},\bm{\sigma}'} \sum_{b_1,\ldots,b_{L-1}}
H_{1,b_1}^{\sigma_1,\sigma_1'} H_{b_1,b_2}^{\sigma_2,\sigma_2'} 
\cdots H_{b_{L-1},1}^{\sigma_L,\sigma_L'}
\ket{\bm{\sigma}} \bra{\bm{\sigma}'} \\
&= \sum_{b_1,\ldots,b_{L-1}} 
\mathcal{H}_{1,b_1} \mathcal{H}_{b_1,b_2} 
\cdots \mathcal{H}_{b_{L-1},1}
\end{aligned}
\label{eq:MPODef}
\end{equation}
with

\begin{equation}
\mathcal{H}_{b_{i-1},b_i} = \sum_{\bm{\sigma}_i,\bm{\sigma}_i'}
H_{b_{i-1},b_i}^{\sigma_i,\sigma_i'} \ket{\bm{\sigma}_i} \bra{\bm{\sigma}_i'}
\label{eq:MPO_Site}
\end{equation}
(note that we dropped any subscript characterizing the Hamiltonian in the two equations above as the MPO decomposition in this form is general).
In Eq.~(\ref{eq:MPODef}), $\mathcal{H}$ is therefore represented as a product of operator-valued matrices $\mathcal{H}_{b_{i-1},b_i}$.
Eq.~(\ref{eq:MPODef}) can be interpreted as the operator counterpart of Eq.~(\ref{eq:MPS_Def}), with the difference that we encode the Hamiltonian exactly as in Eq.~(\ref{eq:MPODef}), while the MPS is an approximation of the exact CI wave function.
Several algorithm to construct MPO representations of operators have been proposed in the literature,\cite{Pirvu2010_MPORepresentation,Frowis2010_MPOGeneric,Hubig2017_GenericMPO,Ren2020}
most of which support only operators with one- and two-body interaction terms.
However, they cannot be applied to the momentum-space representation of the Fermi-Hubbard Hamiltonian of Eq.~(\ref{eq:FH_MomentumSpace}) that contains three-body interactions as well.
We encode such long-range terms in a compact MPO format by generalizing the algorithm applied by us to electronic\cite{Keller2015_MPS-MPO-SQHamiltonian,Baiardi2020_Review} and vibrational\cite{Baiardi2017_VDMRG,Baiardi2019_HighEnergy-vDMRG} Hamiltonians, to three-body interaction terms.
This algorithm starts from a naive MPO representation of the Hamiltonian $\mathcal{H}$, in which the matrices $\mathcal{H}_{b_{i-1},b_i}$ are diagonal, can then compresses it with so-called \quotes{fork} and \quotes{merge} operations.
Without going into the details of the algorithm that can be found in Ref.~\citenum{Keller2015_MPS-MPO-SQHamiltonian}, a fork operation optimizes the representation of the sum of two operators that share the first second-quantized operator is the same (such as, for instance, $a_2^\dagger a_3$ and $a_2^\dagger a_4$).
Similarly, a merge operation optimizes the representation of operators strings for which the last second-quantized operator is the same (such as, for instance, $a_3^\dagger a_4$ and $a_2^\dagger a_4$).
Ref.~\citenum{Keller2015_MPS-MPO-SQHamiltonian} shows that a particularly compact representation of the Hamiltonian is obtained with two fork and one merge compressions.
Following the same idea, we encode three-body terms by applying three fork and two merge compressions.

We highlight that the generality of the algorithm introduced in Refs.~\citenum{Keller2015_MPS-MPO-SQHamiltonian} enables a straightforward support of three-body interaction terms.
This extension would not be as simple within a first-generation DMRG implementation\cite{White1999,Chan2002_DMRG} that constructs the representation of the Hamiltonian from so-called complementary operators, whose definition is limited to two-body interaction terms and would become very complex for three- and higher-body interaction terms.
Even if, as discussed in Ref.~\citenum{Chan2016_MPO-MPS}, the first-generation and MPO/MPS-based formulations of DMRG are formally equivalent, the latter provides a more flexible framework to extend DMRG to transcorrelated Hamiltonians.
Moreover, the Fermi-Hubbard Hamiltonian, both in its real- and momentum-space representations, conserves the overall number of $\alpha$ and $\beta$ electrons.
We exploit this property to construct a symmetry-adapted MPS\cite{Vidal2011_DMRG-U1Symm} in which the $M_{a_{i-1},a_i}^{\sigma_i}$ tensors are block-diagonal and enhance the energy convergence with $m$.

To conclude, we compare the scaling of iTD-DMRG and tcDMRG. In the real-space representation, both Eq.~(\ref{eq:FH_Hamiltonian}) and its transcorrelated counterpart, Eq.~(\ref{eq:FH_Transcorrelated_Real}), contain at most two-body terms. Both operators are therefore represented as MPOs with size scaling as $\mathcal{O}(L^3)$.\cite{Dolfi2014_ALPSProject} However, as we will show in the following section, the energy convergence with $m$ is comparable for both Hamiltonians, and hence, tcDMRG does not lead to any advantage over iTD-DMRG. Conversely, the momentum-space transcorrelated Fermi-Hubbard Hamiltonian includes three-body terms, and therefore the bond dimension of its MPO representation will scale as $\mathcal{O}(L^5)$.
This increases both the computational cost of calculating MPO-MPS contractions and the storage requirements for the boundaries.\cite{Keller2015_MPS-MPO-SQHamiltonian} 
As we will discuss in the next section, this higher computational cost will be balanced by a faster energy convergence with $m$. We highlight that the tcDMRG efficiency can be further enhanced by compressing the MPO representation of the transcorrelated Hamiltonian based on the approach presented in Ref.~\citenum{Frowis2010_MPOGeneric}.
Moreover, all contractions over the $b_i$ MPO indexes can be straightforwardly parallelized as suggested in Refs.~\citenum{Troyer2019_MassivelyParallel-DMRG} and \citenum{Legeza2020_MassivelyParallel-DMRG}. 
The tcDMRG efficiency will strongly benefit from these massively parallelized implementations.

\section{Results}
\label{sec:results}

We applied tcDMRG to the two-dimensional Fermi-Hubbard model for different lattice sizes, fillings, and interaction strengths.
If not otherwise specified, we sorted the orbitals in the one-dimensional DMRG lattice with the so-called snake-like\cite{Legeza2015_Entanglement-2DFermiHubbard} sorting: orbitals corresponding to $i_x$=0 were mapped to the first $W$ sites of the lattice, sorted in increasing $i_y$ values. Then, the orbitals with $i_x=1$ were mapped to sites $W+1,\ldots,2W$ with the same sorting, and this procedure was repeated up to $i_x=L$.
If not otherwise specified, we ran time-independent (TI-)DMRG (usually, we would drop the 'TI' label of this standard version for the sake of brevity, but may keep it here in order to avoid confusion), iTD-DMRG, and tcDMRG calculations with the two-site variant that is less prone to converge into local minima of the energy functional than its single-site counterpart.
All energies are reported in units of the hopping parameter $t$, and all time-steps are expressed in the corresponding reciprocal units.

\begin{figure}[htbp!]
	\centering
	\includegraphics[width=.75\textwidth]{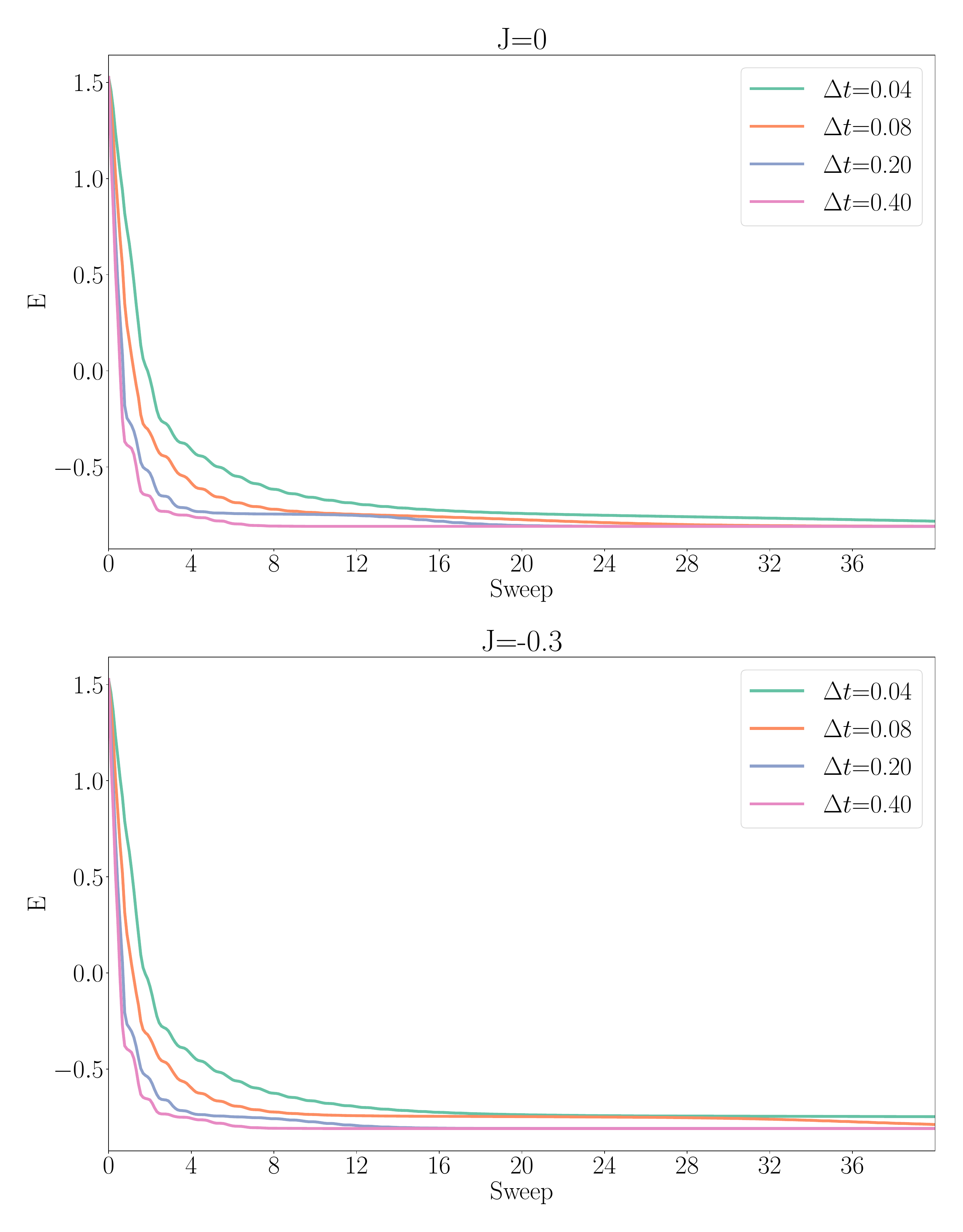}
	\caption{iTD-DMRG (upper panel) and tcDMRG($J$=-0.3) (lower panel) energy convergence with the sweep number for the real-space 3x3 Fermi-Hubbard Hamiltonian for varying $\Delta t$ values. The Hamiltonian parameters are $U$=8, $t$=1, N$_\alpha$=4, N$_\beta$=4, and $m$=300.}
	\label{fig:3x3_Convergence}
\end{figure}

\begin{figure}[htbp!]
	\centering
	\includegraphics[width=.75\textwidth]{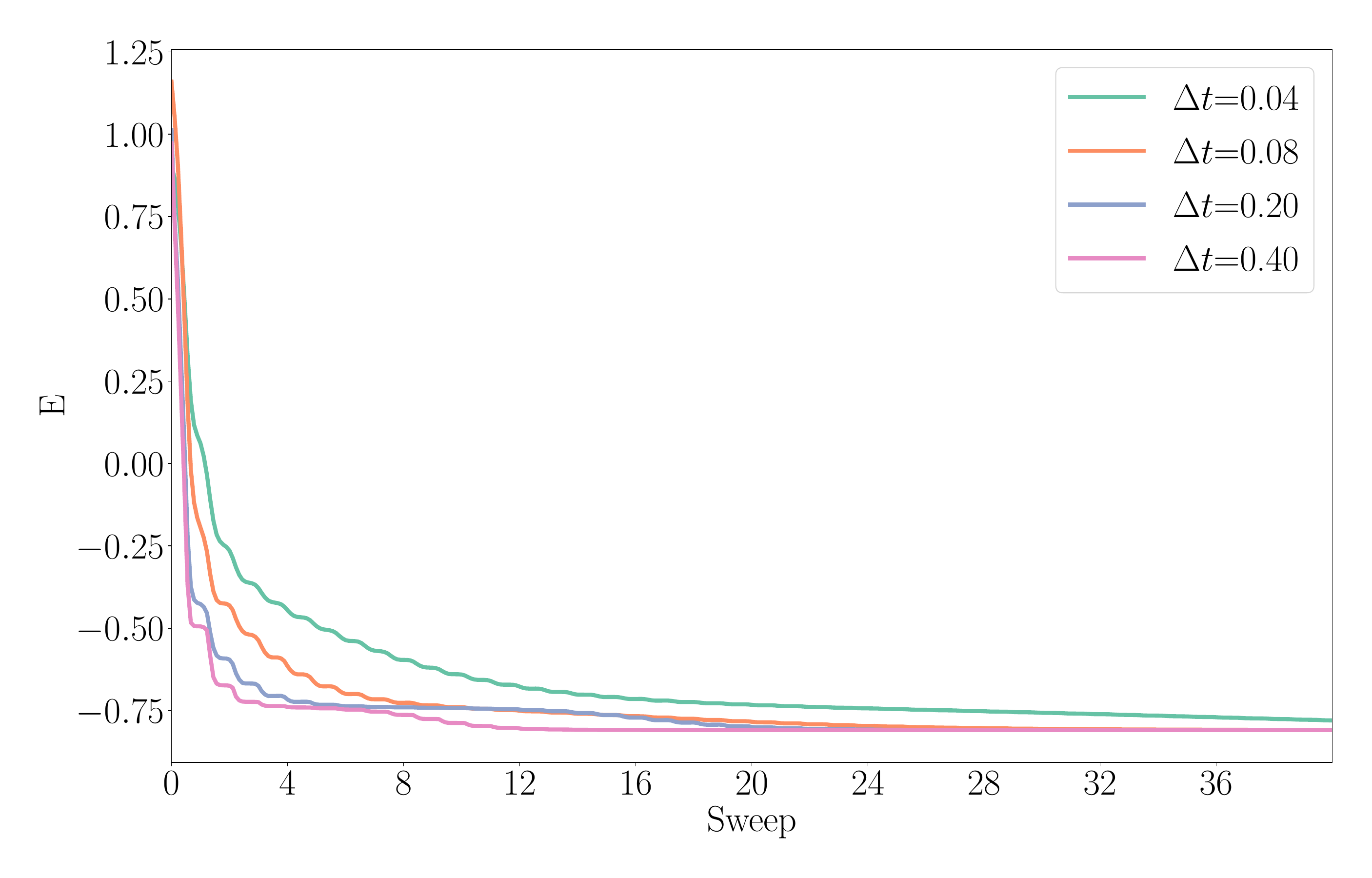}
	\caption{tcDMRG($J$=-0.3) energy convergence with the sweep number for the 3x3 $\bm{k}$-space Fermi-Hubbard Hamiltonian for varying $\Delta t$ values. The Hamiltonian parameters are $U$=8, $t$=1, N$_\alpha$=4, N$_\beta$=4, and $m$=300.}
	\label{fig:3x3_Convergence_Momentum}
\end{figure}

We first analyze the stability of tcDMRG on a 3x3 lattice with $U$=8 and $t$=1, $N_\alpha=4$ and $N_\beta=4$.
We report in Figure~\ref{fig:3x3_Convergence} the energy convergence of iTD-DMRG and tcDMRG($J$=-0.3) for varying time steps for the real-space representation with $m$=300. 
As we will show in the following, this $m$ value delivers converged energies.
Both iTD-DMRG and tcDMRG converge smoothly, the faster convergence being obtained with the largest time step, of 0.40.
For a fixed sweep number, larger time steps corresponds to longer overall propagation times, which enables to reach the $t \rightarrow +\infty$ limit faster.
Note, however, that the computational cost of solving the local differential equation of Eq.~(\ref{eq:LocalDifferentialEquation}) increases with the time step because a larger number of iterations is required to converge the iterative approximation of the exponential operator.
For $\Delta t$=0.40, the Lanczos approximation of the local representation of the imaginary-time propagator converges within 15 iterations in all cases.
These results confirm that, unlike TI-DMRG, imaginary-time TD-DMRG can reliably optimize the ground-state wave function of non-Hermitian Hamiltonians.

As we show in Figure~\ref{fig:3x3_Convergence_Momentum}, the same trend is observed for the $\bm{k}$-space Fermi-Hubbard Hamiltonian.
The fastest convergence (12 sweeps) is observed with $\Delta t$=0.4~au.
With smaller time steps, the energy converges to a local minimum between sweeps 10 and 20, and only afterwards the algorithm converges to the correct limit.
For this reason, if not otherwise stated, in the following we set $\Delta t$=0.40 for all calculations.

\begin{table}[htbp!]
	\begin{tabular}{cc|ccccc}
		\hline \hline
		&  $m$  &  $J$=0   &  $J$=-0.1   &  $J$=-0.3  & $J$=-0.5  &    TI   \\
		\hline
		\multirow{3}{*}{Real space} 
		&  100  & -0.8000  &  -0.8006    &  -0.7999   & -0.7997   & -0.8000 \\
		&  200  & -0.8084  &  -0.8085    &  -0.8084   & -0.8085   & -0.8084 \\
		&  300  & -0.8094  &  -0.8094    &  -0.8094   & -0.8094   & -0.8094 \\
		\hline
		\multirow{3}{*}{$\bm{k}$ space} 
		&  100  & -0.7537  &  -0.7547    &  -0.7616   & -0.7760   & -0.7537 \\
		&  200  & -0.8061  &  -0.8060    &  -0.8063   & -0.8070   & -0.8061 \\
		&  300  & -0.8094  &  -0.8094    &  -0.8094   & -0.8094   & -0.8094 \\
		\hline
		\multirow{3}{*}{$\bm{k}$ space Fiedler} 
		&  100  & -0.7608  &  -0.7620    &  -0.7670   & -0.7770   & -0.7608 \\
		&  200  & -0.8074  &  -0.8074    &  -0.8075   & -0.8082   & -0.8074 \\
		&  300  & -0.8094  &  -0.8094    &  -0.8094   & -0.8094   & -0.8094 \\
		\hline
		\hline
	\end{tabular}
	\caption{Ground-state energy per site of the 3x3 Fermi-Hubbard model obtained with iTD-DMRG and tcDMRG, $U$=8, $t$=1, $N_\alpha$=4, and $N_\beta$=4 for varying bond dimension $m$. The reference energy per site, obtained with exact diagonalization,\cite{Xhang2013_Symmetry-Hubbard-AFQMC} is -0.8094. TI-DMRG energies are reported in the last column.}
	\label{tab:3x3_Results}
\end{table}

We depict in Table~\ref{tab:3x3_Results} the TI-DMRG, iTD-DMRG, and tcDMRG energy convergence with bond dimension $m$ and the transcorrelation parameter $J$ for both Hamiltonian representation.
As expected, all methods converge towards the reference full-CI energy\cite{Xhang2013_Symmetry-Hubbard-AFQMC} with $m$=300.
Note that this $m$ value is large for a 9-orbital system, and this is due to the breakdown of the area law for long-ranged Hamiltonians, as we already noted in Section~\ref{sec:TC_theory}.
As highlighted by Ref.~\citenum{Legeza2015_Entanglement-2DFermiHubbard}, the average interaction range of the Hamiltonian can be strongly reduced by optimizing the orbital sorting in the one-dimensional DMRG lattice based on the so-called Fiedler ordering.\cite{Legeza2011_Entanglement-DifferentStructures}
This orbital sorting minimizes the distance between strongly interacting orbitals, where the interaction strength is evaluated from the mutual information\cite{Legeza2003_OrderingOptimization,Rissler2006_QuantumInformationOrbitals} calculated for the MPS optimized with a partially converged MPS, here obtained with TI-DMRG with $m$=100.
Note that in Ref.~\citenum{Legeza2015_Entanglement-2DFermiHubbard} the impact of the ordering on the wave function entanglement is analyzed only qualitatively.
Here we assess its effect also on the energy convergence with the bond dimension $m$.
As expected, we show in Table~\ref{tab:3x3_Results}, the energy for a given $m$ value is consistently lower with the Fiedler ordering than with the standard one.

We report in Table~\ref{tab:3x3_Results} tcDMRG results obtained with different $J$ values.
For the real-space Fermi-Hubbard Hamiltonian, the tcDMRG energy matches the TI-DMRG one for all $J$ values for $m$=200 and 300.
With $m$=100, the energy difference between TI-DMRG and tcDMRG is smaller than 0.0006 in all cases.
The transcorrelated \textit{ansatz} of Eq.~(\ref{eq:JastrowParametrization}) does not produce therefore a more compact MPS when applied to the real-space two-dimensional Fermi-Hubbard Hamiltonian.
This agrees with the fact that in Ref.~\citenum{Alavi2019_RepulsiveHubbard-Transcorrelated} Alavi and co-workers applied the transcorrelated FCIQMC algorithm to the $\bm{k}$-space two-dimensional Fermi-Hubbard Hamiltonian, and they did not present any results for the real-space representation.

The energy difference between TI-DMRG and tcDMRG is instead considerable for the $\bm{k}$-space representation.
In this case, the difference between the tcDMRG($J$=-0.5) energies obtained with $m$=100 and $m$=300 is nearly halved compared to the corresponding TI-DMRG data.
Note that a correlation factor of $J$=-0.5 is similar to the values optimized in Ref.~\citenum{Alavi2019_RepulsiveHubbard-Transcorrelated} for lattices with similar $U/t$ values.
This confirms that the ground-state wave function of the 3x3 Fermi-Hubbard Hamiltonian can be encoded as a much more compact MPS with the addition of a Gutzwiller correlator. 

\begin{table}[htbp!]
	\begin{tabular}{cc|cccccc}
		\hline \hline
		&  $m$  &  $J$=0   &  $J$=-0.1   &  $J$=-0.3  & $J$=-0.5  &  $J$=-0.5 Fiedler &     TI  \\
		\hline
		\multirow{3}{*}{$\bm{k}$-space} 
		&  500  & -1.0248  &   -1.0249   &  -1.0255   &  -1.0269  &    -1.0260        & -1.0249 \\
		& 1000  & -1.0282  &   -1.0281   &  -1.0284   &  -1.0285  &    -1.0283        & -1.0283 \\
		& 2000  & -1.0288  &   -1.0288   &  -1.0288   &  -1.0288  &    -1.0285        & -1.0288 \\
		\hline
		\hline
	\end{tabular}
	\caption{Ground-state energy per site for the 4x4 Fermi-Hubbard model with $U$=8, $t$=1, $N_\alpha$=4, and $N_\beta$=4. The reference energy, obtained with exact diagonalization,\cite{Xhang2013_Symmetry-Hubbard-AFQMC} is -1.0288.}
	\label{tab:4x4_Quarter}
\end{table}

We report in Table~\ref{tab:4x4_Quarter} the iTD-DMRG and tcDMRG results for a larger, 4x4 lattice with $U$=8, $t$=1, $N_\alpha$=4, and $N_\beta$=4, a parameter set that corresponds to an intermediate correlation regime.
The energy convergence with $m$ is slower than for the 3x3 lattice, and the TI-DMRG energy matches the reference value\cite{Xhang2013_Symmetry-Hubbard-AFQMC} with $m$=2000.
Also in this case, the energy convergence with $m$ is faster for tcDMRG than for iTD-DMRG.
The difference between $m$=500 and fully-converged tcDMRG($J$=-0.5) energies is twice as small than for TI-DMRG.
The lowest energy, for a given $m$ value, is consistently obtained with tcDMRG($J$=-0.5) and the Fiedler orbital sorting.

We have applied so-far tcDMRG to either small or weakly-correlated Hamiltonians, for which iTD-DMRG, when combined with an optimized orbital sorting, can converge the energy with reasonable $m$ values.
The efficiency of tcDMRG becomes apparent in the strong correlation regime, such as for the 4x4 Fermi-Hubbard Hamiltonian with $U$=4, $t$=1, $N_\alpha$=8, and $N_\beta$=8.
We report in Table~\ref{tab:4x4_Half} the corresponding TI-DMRG, iTD-DMRG, and tcDMRG energies for varying $J$ and $m$ values and different orbital sortings.
To avoid convergence into local minima, we adopted the so-called density-matrix perturbation theory approach by White\cite{White2005_DensityMatrixPerturbation} extended to a two-site optimizer for all TI-DMRG simulations.
The computational cost of this perturbative scheme is high, especially for large $m$ values, and would render tcDMRG calculations unpractical.
To avoid a large computational overhead, we adopted the following protocol: we optimized the wave function with TI-DMRG and $m$=500 adding the density-matrix perturbation.
Then, we started TI-DMRG optimizations for larger $m$ values with the resulting MPS as starting guess.
Finally, we ran all tcDMRG calculations for a given $m$ values with the MPS optimized with TI-DMRG and the same $m$ value as initial guess.

\begin{table}[htbp!]
	\begin{tabular}{c|c|cccc}
		\hline \hline
		&  $m$        &  $J$=-0.1  &  $J$=-0.3  & $J$=-0.5  &     TI   \\
		\hline
		\multirow{3}{*}{$\bm{k}$-space} 
		&  500        &  -0.7900   &  -0.7779   &  -0.8496  &  -0.7862 \\
		& 1000        &  -0.8145   &  -0.8279   &  -0.8536  &  -0.8128 \\
		& 2000        &  -0.8310   &  -0.8391   &  -0.8528  &  -0.8297 \\
		\hline
		\multirow{4}{*}{$\bm{k}$-space Fiedler} 
		& 500         &  -0.8485   &  -0.8495   &  -0.8515  &  -0.8484 \\
		& 1000        &  -0.8500   &  -0.8505   &  -0.8513  &  -0.8500 \\
		& 2000        &  -0.8507   &  -0.8511   &  -0.8514  &  -0.8507 \\
		\cline{2-6}
		& 500 Herm.   &  -0.8485   &  -0.8491   &  -0.8496  &  -0.8484 \\
		\hline
		\hline
	\end{tabular}
	\caption{Ground-state energy per site for a 4x4 Fermi-Hubbard model with $U$=4, $t$=1, $N_\alpha$=8, and $N_\beta$=8. The reference energy per site, obtained with exact diagonalization,\cite{Xhang2013_Symmetry-Hubbard-AFQMC} is -0.8514.}
	\label{tab:4x4_Half}
\end{table}

As expected, the energy convergence of TI-DMRG with $m$ is much slower than for the previous Hamiltonian, and the $m$=2000 energy ($-0.8297$) is still far from being converged to the reference result (-0.8514).\cite{Xhang2013_Symmetry-Hubbard-AFQMC}
The energy convergence of tcDMRG with the bond dimension $m$ is, however, much faster.
The faster convergence is delivered by tcDMRG($J$=-0.5) that converges below 0.001 with $m$=2000.

\begin{figure}[htbp!]
	\centering
	\includegraphics[width=.75\textwidth]{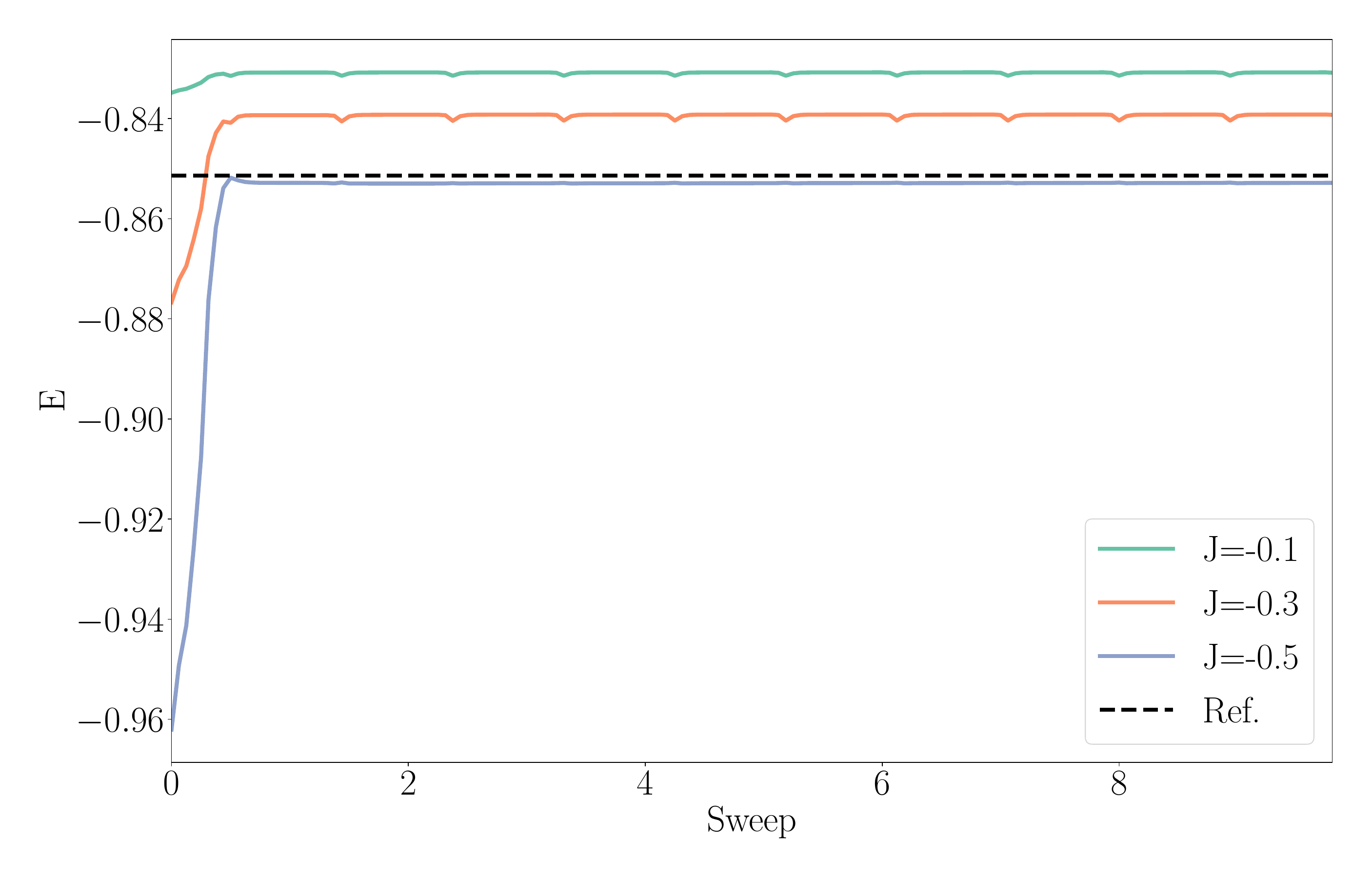}
	\caption{tcDMRG energy convergence for the 4x4 Fermi-Hubbard Hamiltonian with $U/t$=4, $N_\alpha$=8, $N_\beta$=8, for different $J$ values. The initial guess is obtained from the MPS optpimized with TI-DMRG in all cases. The reference energy, taken from Ref.~\citenum{Xhang2013_Symmetry-Hubbard-AFQMC}, is reported as well.}
	\label{fig:ImaginaryTime_Transcorrelated}
\end{figure}

As we show in Figure~\ref{fig:ImaginaryTime_Transcorrelated}, our computational procedure that starts the tcDMRG imaginary-time propagation from the MPS optimized with TI-DMRG converges the energy efficiently with 2-3 sweeps.
It is also worth noting that the energy is lower than the reference, full-CI value for $J$=0.3 and $J$=0.5.
We recall that the energy is the expectation value of the transcorrelated Hamiltonian (see Eq.~(\ref{eq:FH_Transcorrelated_Momentum})) over the MPS and, as we already highlighted in Section~\ref{sec:TC_theory}, the variational principle does not apply.
Therefore, an energy lower than the exact, full-CI one is physically acceptable in this case.
We highlight that, unlike tcDMRG, a variational-based optimization starting from the TI guess would not converge to the correct minimum.
We conclude by noting that the $J$ value that provides the best match with the reference data is 0.5.
This value agrees with the optimal value obtained in Ref.~\citenum{Alavi2019_RepulsiveHubbard-Transcorrelated} based on the optimization strategy described in Ref.~\citenum{Scuseria2015_LieAlgebraic} for a 18-sites Fermi-Hubbard Hamiltonian with the same $U/t$ value.
This suggest that the same algorithm,\cite{Scuseria2015_LieAlgebraic} that requires the solution of a Coupled Cluster-like equation, can be effectively applied to determine the optimal $J$ value for tcDMRG.
Note that the results reported in Figure~\ref{fig:ImaginaryTime_Transcorrelated} also indicate that the transcorrelation parameter $J$ cannot be optimized variationally.

As we show in Table~\ref{tab:4x4_Half}, the orbital sorting has a critical impact in the energy convergence of DMRG.
With the Fiedler ordering, the TI-DMRG energy obtained with $m$=500 is only 0.0023 higher than the reference energy, while the same difference with the Fiedler ordering is larger than 0.04.
However, even by adopting the Fiedler ordering, the energy is not converged even with $m$=2000.
By combining the optimized orbital ordering with tcDMRG($J$=-0.5), the energy differs from the reference one by only 10$^{-4}$ already with $m$=500.
Therefore, the energy convergence of tcDMRG with the bond dimension is in this case truly faster than that of TI-DMRG.
This further confirms that, also in the presence of strong correlation, the ground state of the Fermi-Hubbard model can be parametrized as Eq.~(\ref{eq:JastrowParametrization}), where $\ket{\Phi}$ can be efficiently represented as a low-entanglement wave function and optimized by applying iTD-DMRG to the transcorrelated Hamiltonian.

Alavi and co-workers\cite{Alavi2019_RepulsiveHubbard-Transcorrelated} showed that, if, the right lowest-energy eigenvector of the transcorrelated Hamiltonian can be represented as compact CI wave functions, the left one will be represented by a much less compact expansion.
Similarly, as we show in the last row of Table~\ref{tab:4x4_Half}, the convergence of the energy of the left eigenvector is slower than for the right one (as discussed in Ref.~\citenum{Alavi2019_RepulsiveHubbard-Transcorrelated}, the right-eigenvector corresponding to a given $J$ value can be optimized with tcDMRG by setting the Jastrow factor to $-J$).

\begin{figure}[htbp!]
	\centering
	\includegraphics[width=.75\textwidth]{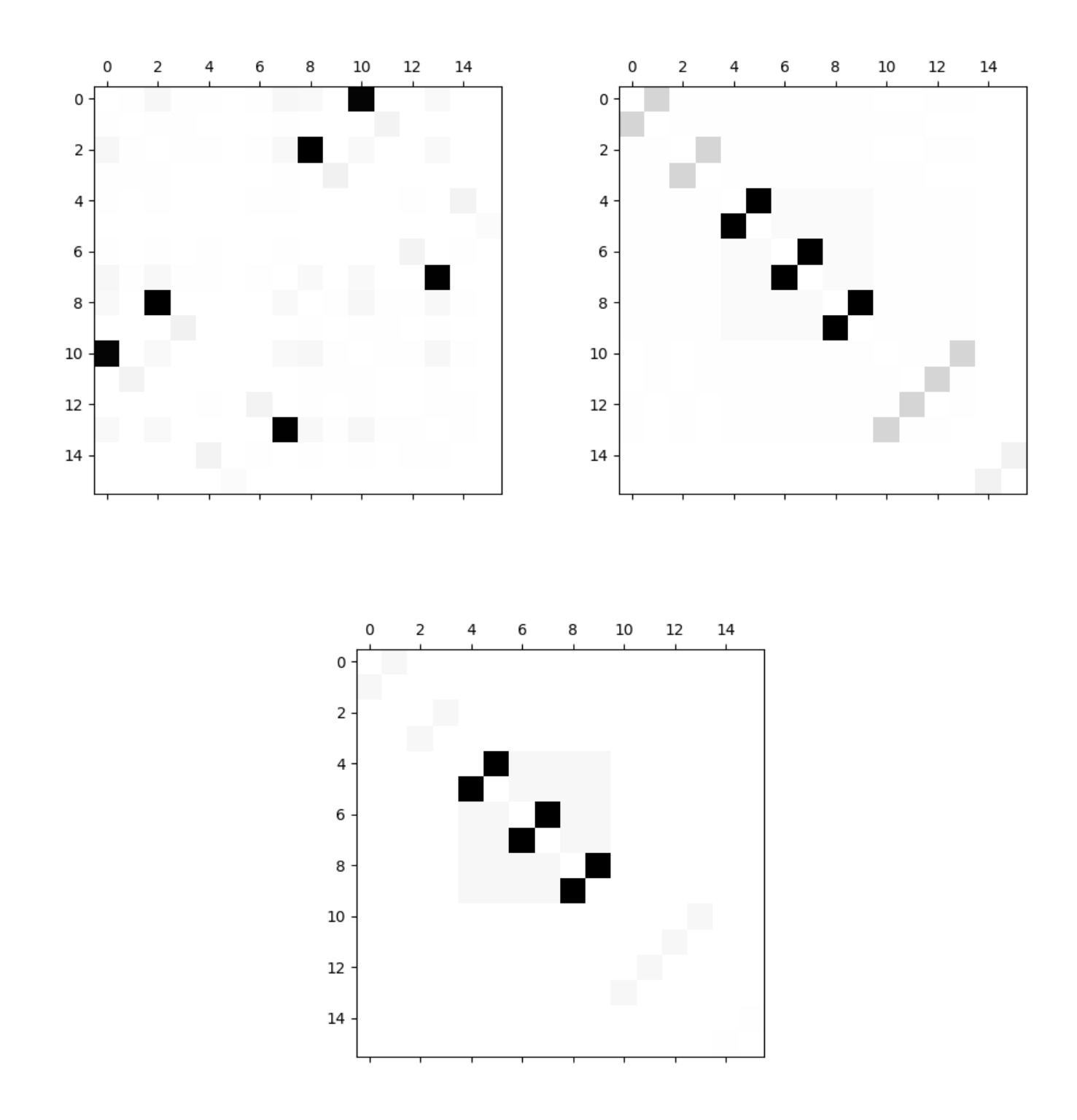}
	\caption{Graphical representation of the mutual information matrix $I_{ij}$ for a 4x4 Fermi-Hubbard lattice with $U/t$=4 at half filling, obtained with TI-DMRG based on the standard ordering (upper left), TI-DMRG with the Fiedler ordering (upper right), and tcDMRG with the Fiedler ordering (bottom).}
	\label{fig:EntanglementStructure}
\end{figure}

To further characterize the entanglement structure of the ground-state wave function of the transcorrelated Hamiltonian, we report in Figure~\ref{fig:EntanglementStructure} a graphical representation of the orbital mutual information matrix $I_{ij}$,\cite{Legeza2003_OrderingOptimization,Rissler2006_QuantumInformationOrbitals,Stein2019_AutoCAS-Implementation} with 

\begin{equation}
  I_{ij} = \frac{1}{2} \left[ s_i(1) + s_j(1) - s_{ij}(2) \right] (1 - \delta_{ij}) \,
  \label{eq:MutualInfo}
\end{equation}

calculated with TI-DMRG and tcDMRG($J$=-0.5) for different orbital orderings.
$s_i(1)$ is the single-orbital entropy for orbital $i$, defined as

\begin{equation}
s_i(1) = -\sum_{\alpha=1}^4 w_{i,\alpha} \ln w_{i,\alpha} \,
\label{eq:SingleOrbitalEntropy}
\end{equation}
where $w_{i,\alpha}$ is the $\alpha$-th eigenvalue of the one-orbital reduced density matrix.
Similarly, $s_{ij}(2)$ is the two-orbital entropy, defined as

\begin{equation}
s_{ij}(2) = -\sum_{\alpha=1}^{16} w_{ij,\alpha} \ln w_{ij,\alpha} \,
\label{eq:TwoOrbitalEntropy}
\end{equation}
where $w_{ij,\alpha}$ the $\alpha$-th eigenvalue of the two-orbital reduced density matrix for orbitals $i$ and $j$.
As expected, the mutual information matrix has a sparse structure with the standard ordering, while it becomes diagonally dominant for the MPS constructed with the Fiedler ordering.
Most importantly, the magnitude of most non-zero elements of the TI-DMRG mutual information matrix are much smaller for tcDMRG($J$=-0.5).
This further confirms that the multi-reference character, measured as orbital entanglement, of the ground state of the transcorrelated Hamiltonian is much smaller than that of the original Fermi-Hubbard Hamiltonian. 
For this reason, the former can be much more efficiently represented as an MPS.
We recall that a similar regularization of the orbital mutual information has been previously observed when combining DMRG with short-range DFT (DMRG-srDFT).\cite{Hedegard2015_DMRG-srDFT}
In DMRG-srDFT the Hamiltonian is modified to include only the long-range part of the electron-electron Coulomb interaction.
Similarly, tcDMRG removes dynamical correlation with the Jastrow factorization of the wave function.
In both cases, the mutual information associated to the DMRG wave function becomes more sparse since it is large only for the orbitals coupled by pure static correlation effects.

\begin{table}[htbp!]
	\begin{tabular}{cccc|cc|c}
		\hline \hline
		$U/t$ & N$_\alpha$ & N$_\beta$ &  $m$ &  TI-DMRG   & tcDMRG  &    Ref.   \\
		\hline
		\hline
		4    &    12      &     12    &  500 &  -1.1500   & -1.1804  &  -1.1853  \\
		2    &    18      &     18    &  500 &  -1.1345   & -1.1530  &  -1.1516  \\
		4    &    18      &     18    &  500 &  -0.8206   & -0.8596  &  -0.8574  \\
		4    &    18      &     18    & 1000 &  -0.8307   & -0.8580  &  -0.8574  \\
		\hline
		\hline
	\end{tabular}
	\caption{TI-DMRG and tcDMRG ground-state energy per site of a 6x6 Fermi-Hubbard model with different number of electrons and interaction strengths, for varying $m$ values. The reference benchmark data are taken from Refs.~\citenum{Xhang2013_Symmetry-Hubbard-AFQMC} and \citenum{Qin2016_2DHubbardBenchmark} and are calculated with auxiliary field Quantum Monte Carlo.}
	\label{tab:6x6}
\end{table}

We report in Table~\ref{tab:6x6} the TI-DMRG and tcDMRG energies of the 6x6 Fermi-Hubbard lattice, a system that has been studied with the transcorrelated variant of FCIQMC\cite{Alavi2019_RepulsiveHubbard-Transcorrelated} and which is out of the reach for exact-diagonalization approaches, for various fillings and $U/t$ ratios.
To limit the computational cost, we obtained the tcDMRG results with the single-site imaginary-time propagator starting from the MPS optimized with TI-DMRG.
In all cases, we sorted the orbitals in the lattice with the Fiedler ordering obtained with TI-DMRG($m$=500).
Moreover, we set the parameter $J$ to the optimal value taken from Ref.~\citenum{Alavi2019_RepulsiveHubbard-Transcorrelated}.
The results reported in Table~\ref{tab:6x6} confirm that, also for large lattices, the energy convergence with $m$ of tcDMRG is much faster than for standard TI-DMRG.
In all cases, the difference between TI-DMRG($m$=500) and the reference energies is larger than 0.02. 
The error becomes one order of magnitude smaller with tcDMRG($m$=500), and is consistently lower than 0.004.
By further increasing $m$ to 1000, the energy of the half-filled 6x6 lattice with $U/t$=4, which is the most strongly correlated lattice among the ones studied here, the tcDMRG energy matches the reference one with an error smaller than 10$^{-3}$.

\section{Conclusions}
\label{sec:conclusions}

In this work, we introduced transcorrelated DMRG theory that optimizes the ground-state wave function of the transcorrelated Hamiltonian as a matrix product state.
The optimization algorithm is tailored to the problem: imaginary-time time-dependent DMRG. Unlike standard time-independent DMRG relying on the variational principle, iTD-DMRG can reliably optimize the ground-state energy of non-Hermitian Hamiltonians.
We applied tcDMRG to the two-dimensional Fermi-Hubbard Hamiltonian for different sizes, fillings, and interaction strengths.
We demonstrate that, both for weak and strong correlation regimes, the energy convergence of tcDMRG is consistently much faster than that of TI-DMRG.
In practice, tcDMRG can converge the energy of Fermi-Hubbard lattices including up to 36 sites, which otherwise would be a challenge for standard DMRG approaches due to the constraints imposed by the area law.
Our results agree with recent findings in the framework of FCIQMC by Alavi and co-workers, who showed that the ground-state of the electronic\cite{Alavi2019_Transcorrelated-Molecules} and Fermi-Hubbard\cite{Alavi2019_RepulsiveHubbard-Transcorrelated} transcorrelated Hamiltonians are efficiently represented by very compact configuration-interaction expansions.
In addition, we showed in the present work that the ground-state wave functions belong, more generally, to the class of low-entanglement wave functions and can be therefore encoded as a compact MPS.
Note that this conclusion extends the findings of Alavi and co-workers\cite{Alavi2019_RepulsiveHubbard-Transcorrelated} since a compact MPS wave function can encode both sparse and dense full-CI expansions.

The successful application of the tcDMRG algorithm to the two-dimensional Fermi-Hubbard Hamiltonian suggests that the same theory can be applied as successfully to the electronic Hamiltonian as well.\cite{Alavi2019_Transcorrelated-Molecules}
In this case, the Jastrow factor in conveniently expressed in real space because the similarity-transformed Hamiltonian includes exactly only up to three-body terms, the main hurdle being the need of calculating the resulting three-centers integrals.
However, we expect that tcDMRG will be even more efficient when applied to electronic-structure problems for two reasons.
In fact, the Gutzwiller correlator of the transcorrelated Fermi-Hubbard Hamiltonian is governed by a single parameter.
Conversely, the real-space Jastrow factor includes a different correlation parameter for each orbital pair.
The wave function parametrization is, therefore, much more flexible and can be adapted to the specific molecular system under analysis.
Moreover, the three-body potential term appearing in the transcorrelated Fermi-Hubbard Hamiltonian couples any possible combination of six different orbitals.
The MPO representation of the resulting Hamiltonian is highly non-compact, even though in the present work we discussed a way to encode it efficiently.
For the quantum-chemical case, it will be possible to exploit orbital locality to screen and compress the three-body part of the Hamiltonian and encode it as a compact MPO and to further enhance the tcDMRG efficiency.
We showed that the similarity transformation underlying tcDMRG regularizes the wave function entanglement and makes the distinction between strongly- and weakly-correlated orbitals more rigid.
This suggests that tcDMRG can be effectively combined with active-space based approaches, where dynamical correlation effects are added \textit{a-posteriori} with perturbation theory or with multi-reference variants of the Coupled-Cluster method.\cite{Paldus1997_ExternallyCorrected,Bartlett2005_TCC,Veis2016_TCC-DMRG,Evangelista2018_MRCC-Review}
All these extensions are currently explored in our laboratory and results will be reported in future work.

\begin{acknowledgements}
This work was supported by ETH Z\"{u}rich through the ETH Fellowship No. FEL-49 18-1.
\end{acknowledgements}

\end{document}